# Impact of Motor Stator Winding Faults on Common-Mode Current

Fei Fan, Zhenyu Zhao, Pengfei Tu, Huamin Jie, Kye Yak See
School of Electrical and Electronic Engineering
Nanyang Technological University
Singapore
fanf0003@e.ntu.edu.sg

*Abstract*—This paper investigates the influence of different motor stator failures on the common-mode (CM) current. The stator winding failures will affect the motor's CM impedance, and also increase the unbalance of the differential-mode (DM) noise path. The experimental results show that the former will increase the CM current induced by the CM noise source, while the latter will cause the DM current induced by the DM noise source to be converted into the CM current.

## I. Introduction

Performance improvements and declining prices of the fast switching power electronics devices have gained popularity in the adoption of variable frequency drives (VFDs) for motor drive systems. However, the high-speed switching process also causes significant common-mode (CM) voltage with high dv/dt [1]. The resultant CM current at the output of the VFD can lead to electromagnetic interference (EMI) problems and motor bearing failures [2]. The characteristics of the CM current and its mitigation techniques for a motor drive system have been well studied and reported [3], [4]. However, for a motor with incipient stator winding faults, the influence of these faults on the CM current has not been well studied and evaluated.

As one of the most prevalent modes of motor faults, the stator winding fault accounts for nearly 40 % of the total faults of the induction motors [5]. The stator winding fault is usually caused by turn-to-turn, phase-to-phase and phase-to-ground shorts, as illustrated in Fig. 1. In a highly reliable Insulated Terrestrial (IT) system, the operation of the motor drive system will not be affected when the above-mentioned stator winding shorts happen in the initial state [6] but if left unattended, they will lead to irreversible damage to the motor. The shorts will cause imbalance of the differential-mode (DM) current path of the motor, which can influence the resultant CM current. Unlike the DM current, the CM current of the machine is usually stable as the operation condition does not affect the stray capacitance between the stator winding and the machine frame. Hence, this paper explores the use of CM current changes as a monitoring parameter for the early detection of incipient stator winding fault. Using a 5.5-kW induction motor as a test case, the stator windings of the motor are modified to emulate the various types of winding shorts so that the impact of their influence on the CM current can be studied and analysed.

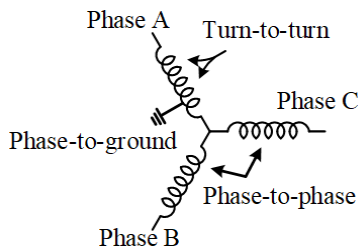

Fig. 1. Prevalent motor faults due to stator winding shorts.

## II. Equipment Under Test

A three-phase induction motor (TECO 1071033064C-1, 5.5 kW, 400 V, 6-pole and 960 rpm) is chosen for the study in this paper. The three-phase stator windings are in star-configuration and there are in total 288 turns of winding per phase. The emulation of the stator winding fault is implemented by tapping the respective points of the windings and the motor frame, as indicated in Fig. 2. The turn numbers 24, 27, 34, 120 and 264 are tapped from Phase-A and the turn number 120 is tapped from Phase-B. To emulate the turn-to-turn fault of different levels of severity, two adjacent tapped points in Phase-A are shorted by a 1-Ω resistor [7], i.e. $A_{24}$-$A_{27}$ and $A_{24}$-$A_{34}$. Similarly, the phase-to-phase faults are emulated by shorting $A_{120}$-$B_{120}$ and $A_{264}$-$B_{120}$ while the phase-to-ground fault is emulated by shorting $A_{24}$-G.

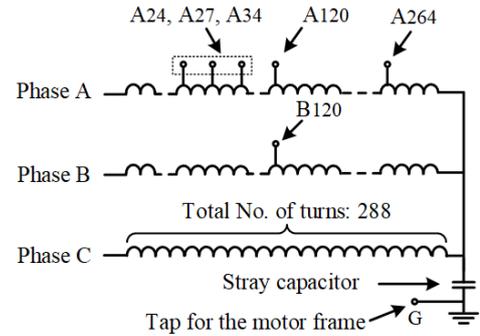

Fig. 2. Tapping points of the 3-phase stator winding for fault emulation.

## III. Influence on the CM Current

A stator winding fault will lead to a low contact impedance between the shorted turns or phases and thus introduce imbalance of the three-phase windings. This imbalance results in changes of the CM impedance of the motor and the resultant CM current. Fig. 3(a) shows a typical three-phase motor drive system. Fig. 3(b) shows its equivalent CM circuit model, where the VFD and the induction motor are modelled as a CM noise source and a CM noise termination, respectively. In the equivalent circuit, $V_{VFD,CM}$ and $Z_{VFD,CM}$ are the equivalent CM noise voltage source and noise source impedance of the VFD, respectively; $Z_{W,CM}$ and $Z_{IM,CM}$ are the equivalent CM impedance of the three-phase cable and induction motor, respectively. With the equivalent circuit model, the CM current in the motor drive system can be determined by:

$$I_{CM} = \frac{V_{VFD,CM}}{Z_{VFD,CM} + Z_{W,CM} + Z_{IM,CM}} \quad (1)$$

To evaluate the influence on the CM current, a parameter $R$ is defined as the ratio of the CM current $I'_{CM}$ with a stator winding fault and the CM current $I_{CM}$ of a healthy stator winding (expressed in dB):

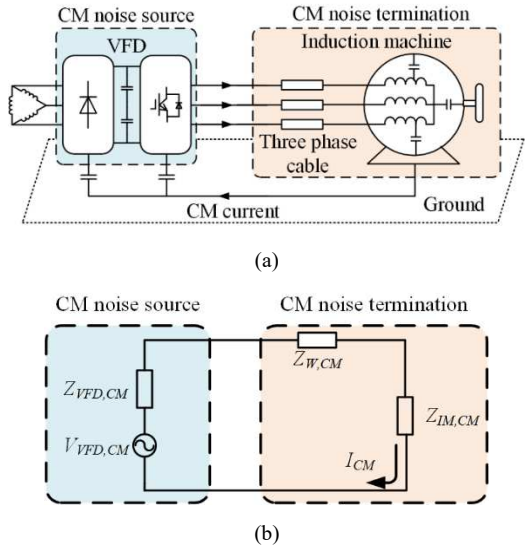

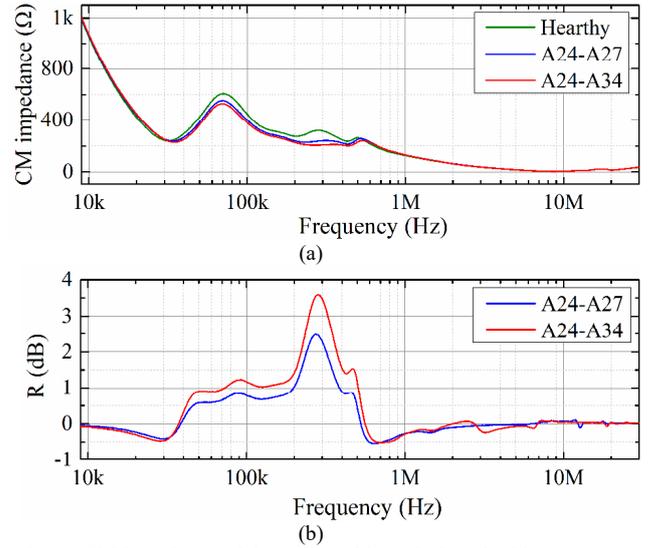

(a)

(b)

Fig. 3. (a) CM current path of a motor drive system (b) CM equivalent circuit model.

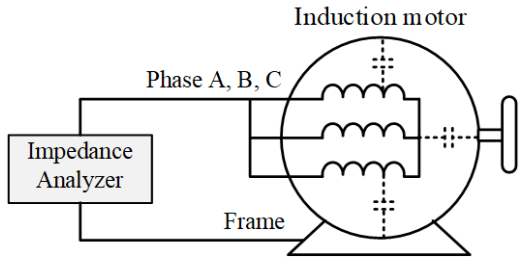

Fig. 4. Measurement of the motor's CM impedance with an impedance analyzer.

$$R = 20\log\left|\frac{I'_{CM}}{I_{CM}}\right| = 20\log\left|\frac{Z_{VFD,CM} + Z_{W,CM} + Z_{IM,CM}}{Z_{VFD,CM} + Z_{W,CM} + Z'_{IM,CM}}\right| \quad (2)$$

where $Z'_{IM,CM}$ is the measured CM impedance of the motor with respective emulated stator winding faults.

For a healthy motor without any faults, the CM impedance is mainly contributed by the stray capacitance between the stator winding and the motor frame, which is usually very stable. However, under various stator winding faults, the CM impedance can vary due to the imbalance introduced by the faulty stator winding. Prior to the actual online measurement, the motor's CM impedance is measured offline using an impedance analyzer (OMICRON Lab Bode 100), as shown in Fig. 4 [8].

Fig. 5(a) compares the measured CM impedance of the motor without fault and with the turn-to-turn faults (i.e., $A_{24}$-$A_{27}$ short and $A_{24}$-$A_{34}$ short). It is clearly observed that the deviation of the CM impedance for the fault $A_{24}$-$A_{34}$ is more visible due to more number of turns shorted [9]. The online noise source impedance and the cable impedance for the motor drive system have been extracted in [10], which can be substituted into (2) to calculate the $R$ of the CM currents, as plotted in Fig. 5(b). For a 10-turn short, $R$ has an increase of 3.5 dB at 280 kHz, which can cause the resultant CM current to increase and pose an EMI problem, in addition to the fault.

Another more serious fault that can occur in the stator windings is the phase-to-phase short. The higher voltage difference between phases can accelerate the fault to an irreversible damage of the motor. However, if the phase-to-phase fault occurs near the neutral point of a star-connected motor, this may not affect the short-term operation of the motor, but it is good to detect it early for the necessary preventive measures. The phase-to-phase faults $A_{120}$-$B_{120}$ and $A_{264}$-$B_{120}$ shorts are emulated to study their influence on the CM current. The resultant CM impedances of the motor and the corresponding variation of the CM currents are plotted in Figs. 6(a) and (b), respectively. The $A_{264}$-$B_{120}$ short causes a shift of the resonant frequency of the stator windings and enhances the CM current at around 60 kHz. On the contrary, the $A_{120}$-$B_{120}$ short has a limited impact on the CM impedance, which is expected because $A_{120}$ and $B_{120}$ are at two equipotential points from the CM analysis perspective.

Fig. 5. (a) CM impedance of the motor without fault and with turn-to-turn faults $A_{24}$-$A_{27}$ and $A_{24}$-$A_{34}$; (b) Variation of the CM currents.

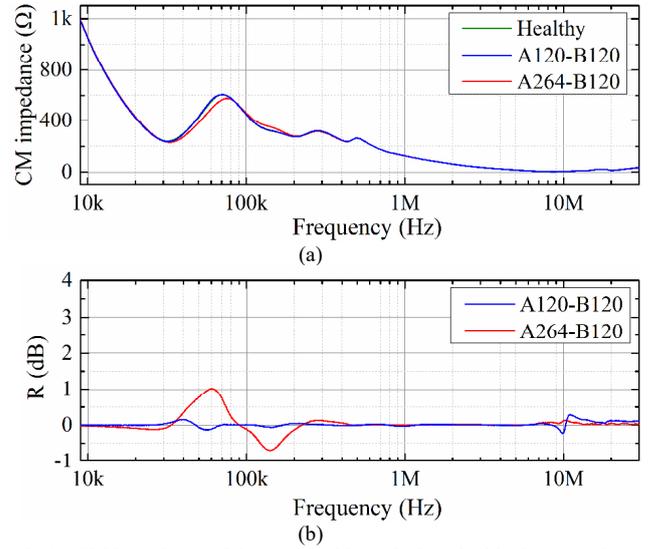

Fig. 6. (a) CM impedance of the motor without fault and with phase-to-phase faults A120-B120 and A264-B120; (b) Variation of the CM currents.

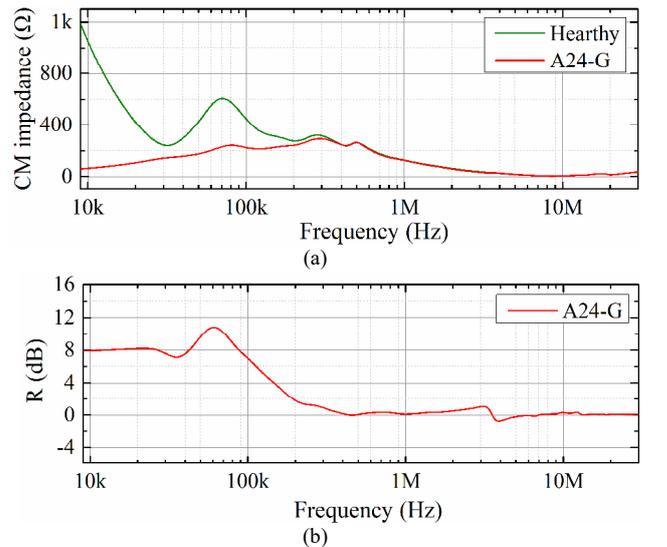

Fig. 7. (a) CM impedance of the motor without fault and with phase-to-ground faults A24-G; (b) Variation of the CM current.

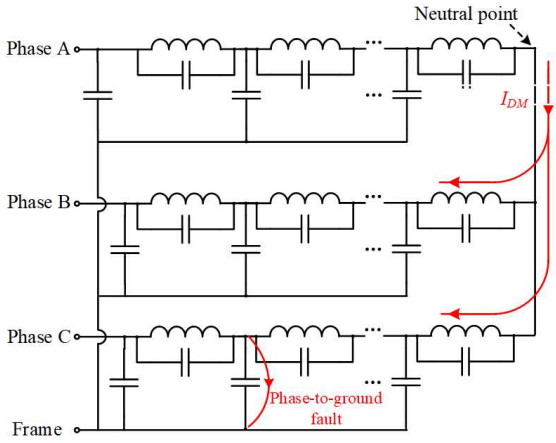

Fig. 8. Distributed equivalent-circuit model of the induction motor

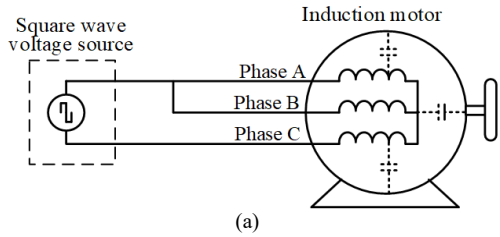

(a)

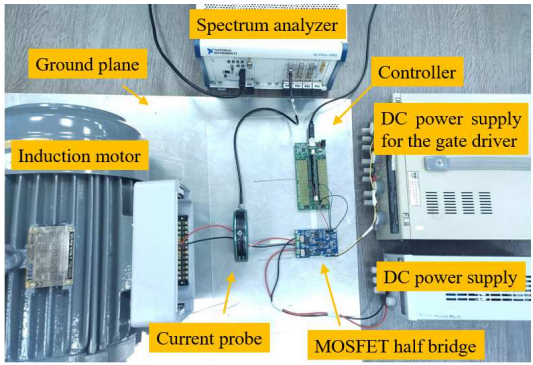

(b)

Fig. 9. (a) Test setup to evaluate the DM-to-CM conversion caused by the stator winding failures (b) Actual photo.

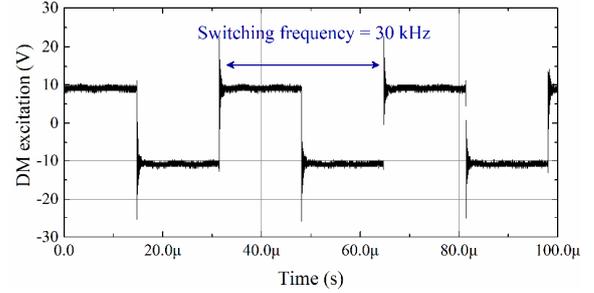

Fig. 10. Waveform of the DM excitation.

In a Terre-Neutral (TN) or Terre-Terre (TT) system, the neutral is always grounded and thus the ground fault relay trips immediately should a phase-to-ground fault occurs. However, such a fault in a highly reliable IT system will not result in high enough fault current because the IT system is configured with the three phases at live voltages all floating and there is no closed loop for a line-frequency current to flow. Therefore, the EMI-related issue caused by the motor's phase-to-ground faults is worth investigating in the scenario of IT systems. Fig. 7(a) shows the CM impedances of the motor without fault and with the $A_{24}$-G short. The motor's CM impedance without fault is capacitive in nature at low frequency because of the dominant stray capacitance between the stator winding and the motor frame. The phase-to-ground fault creates a low impedance path in parallel with this stray capacitance, thus the motor's CM impedance is reduced significantly. With the increasing frequency, the equivalent series inductance in the CM path becomes dominant, and consequently, its low impedance weakens the effect of the phase-to-ground fault on the CM impedance. This can also be reflected from the variation of the calculated CM current in Fig. 7(b), where a significant increase of CM current is observed only below 400 kHz.

## IV. DM-TO-CM CONVERSION

The stator winding faults not only affect the motor's CM impedance but also cause the unbalance of the DM noise current path which could lead to the transformation from the DM noise to the CM noise [10]. It is reported in [11] that the CM current transformed from the high-amplitude DM noise could be even higher than the original CM current. Also, the mode transformation is usually not considered when designing the EMI filter. Therefore, it is worthwhile to investigate the DM-to-CM conversion caused by the stator winding faults and its influence on the CM current.

Fig. 8 shows the distributed equivalent-circuit model of the induction motor where a phase-to-ground fault is used as an example. With the additional current path, part of the DM current $I_{DM}$ is diverted from Phase C to the ground, which is coincides with the CM current path. This equivalent circuit can be used to simulate the mode conversion by emulating a short circuit at a specific fault location. However, it requires detailed geometric information of the motor to predetermine the distributed parameters with Finite Element Method (FEM) analysis [12]. This paper evaluates the mode conversion experimentally under various emulated faults, which eliminates the needs to know the detailed information of the motor.

With the setup in Figs. 9(a) and (b), a purely DM square wave is applied to the induction motor by shorting phase A and phase B together. The 30 kHz square wave has a voltage swing of ±10V, which is generated using SiC MOSFET switches and a digital signal processor to emulate the PWM voltage from the VFD [13], as illustrated in Fig. 10. The high dv/dt results in a transient voltage spike up to 25 V due to the the parasitic elements. The CM current flows through the stray capacitance between the voltage source and the ground plane and is monitored by a current probe (Pearson 6600) and an oscilloscope (NI PXI-5922) with spectrum analyzer function, which should be zero for an ideal balanced system since there is no DM-to-CM conversion happened in the circuit [14].

Figs. 11, 12 and 13 show the CM currents induced by the DM excitation for the turn-to-turn, phase-to-phase and phase-to-ground faults, respectively. The x-axis is expressed in terms of the harmonic number for better demonstration of the mode

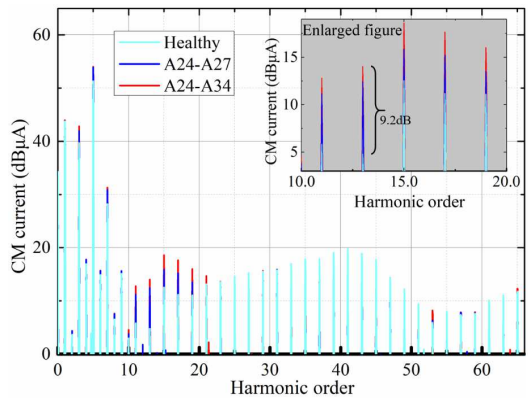

Fig. 11. CM currents without fault and with turn-to-turn faults A24-A27 and A24-A34 under the DM excitation (fundamental frequency = 30 kHz).

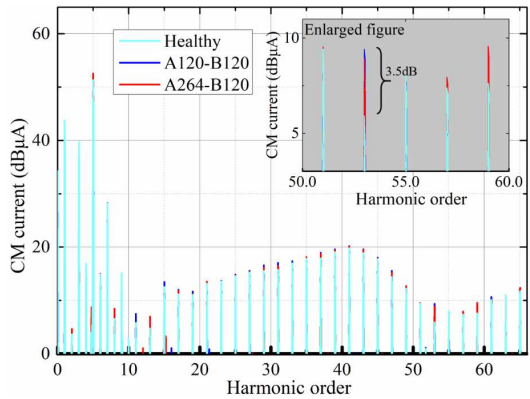

Fig. 12. CM currents without fault and with phase-to-phase faults A120-B120 and A264-B120 under the DM excitation (fundamental frequency = 30kHz).

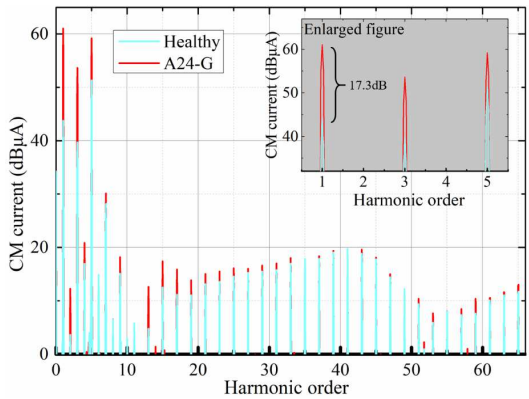

Fig. 13. CM currents without fault and with turn-to-ground fault A24-G under the DM excitation (fundamental frequency = 30kHz).

conversion. In reality, finite CM current exists even without any stator winding faults because there is always an inherent asymmetry in the electrical motor drive system due to some tolerances in the manufacturing process [15]. Nevertheless, the CM current measured at the no-fault condition can be used as a reference to evaluate the DM-to-CM conversion caused by the above-mentioned motor faults. It is clearly observed that for all three types of faults, there is an increasing trend of CM current, which verifies the existence of the DM-to-CM conversion for a faulty motor. As highlighted in the enlarged images in Figs. 11, 12 and 13, the most prominent increments in the CM currents are 9.2 dB, 3.5 dB and 17.3 dB for the emulated turn-to-turn, phase-to-phase and phase-to-ground faults, respectively.

## V. CONCLUSION

The stator winding faults account for a significant portion of the failures in a motor drive system. Before the complete motor failure, the motor with an incipient stator winding fault can increase the CM current significantly and cause potential EMI related issues. In this paper, it has been demonstrated experimentally that the change of the CM impedance and the imbalance introduced by the stator winding faults have a significant impact on the characteristics and level of the CM current. It also reveals the feasibility of fault detection by monitoring the CM current. The future work will be a detailed FEM analysis for an induction motor to quantify the DM-to-CM conversion with various fault types and severities.